Comments on «Thermic and Caloric Equations of State with Small Number of Parameters»

Umirzakov I.H.

*E-mail: tepliza@academ.org*

ABSTRACT

It is shown that the potential of Keesom which depends on temperature cannot be used for calculate  the second virial coefficient of polar molecules and the formulae for second virial coefficient of [2]  have no molecular statistical mechanical base for water and carbon dioxide. The contradictions and errors of [2] are discussed.

## КОММЕНТАРИИ К «МАЛОПАРАМЕТРИЧЕСКИЕ ТЕРМИЧЕСКИЕ И КАЛОРИЧЕСКИЕ УРАВНЕНИЯ СОСТОЯНИЯ РЕАЛЬНОГО ГАЗА»

И.Х. Умирзаков

*E-mail: tepliza@academ.org*

### АННОТАЦИЯ

Показано, что нельзя использовать зависящий от температуры потенциал Кеезома для описания второго вириального коэффициента полярных молекул вместо потенциала взаимодействия электрических диполей,   работа [2] внутренне противоречива, второй вириальный коэффициент и уравнения состояния в ней не имеют молекулярного статистическо-механического обоснования для воды и диоксида углерода.  Рассмотрены некоторые противоречия и ошибки, допущенные в этой статье.

### ВВЕДЕНИЕ

Зависящие от температуры потенциалы для практических приложений  были получены Кеезомом путем усреднения по всем ориентациям молекул зависящих от взаимной ориентации молекул электростатических потенциалов взаимодействия постоянных электрических мультиполей молекул [1]. В работе показано, что потенциал Кеезома нельзя использовать для определения ВВК. Обсуждаются некоторые внутренние противоречия и ошибки, имеющиеся в статье [2].

# РЕЗУЛЬТАТЫ

1.    Известно, что второй вириальный коэффициент (ВВК) *B(T)* для не зависящего от взаимной ориентации молекул потенциала взаимодействия между молекулами *U(r)* может быть вычислен по формуле

$$B(T) = 2\pi \int_0^\infty (1 - e^{-U(r)/kT}) r^2 dr. \qquad (1)$$

Как видно из формулы (15.1) на стр.756 в [1] параметр $\alpha$ потенциала Кеезома $\varphi_K = -\alpha / kTr^6$, используемого в работе [2], равен $\alpha = 2\mu^4 / 3$, где $\mu$ - электрический дипольный момент молекулы (мы используем обозначения работы [2]).

Легко показать, что если ограничиться только первыми двумя членами разложения в ряд Тейлора экспоненты под интегралом в (1), использовать суммы потенциалов взаимодействия Лондона и Кеезома для *r>d*, как это сделано в [2], то отвечающий диполь-дипольному взаимодействию и зависящий от квадрата обратной температуры член в разложении ВВК ( $D = d$ в [2]), равен

$$-\frac{2\pi}{3} D^3 N_A \cdot \frac{2}{3} \cdot \left( \frac{\mu^2}{D^3 kT} \right)^2. \qquad (2)$$

Как видно из формулы (10.1) на стр.177 в [1] зависящий от квадрата обратной температуры член в разложении ВВК твердых сфер радиуса *D* ( $D = \sigma$ в [1]), содержащих точечные электрические диполи, равен

$$-\frac{2\pi}{3} D^3 N_A \cdot \frac{1}{3} \cdot \left( \frac{\mu^2}{D^3 kT} \right)^2. \qquad (3)$$

Из сравнения (2) и (3) видно, что абсолютное значение первого ненулевого вклада от диполь- дипольного взаимодействия в ВВК в работе [2] в два раза больше чем абсолютное значение правильного вклада (3).

Можно показать, что использование зависящих от температуры потенциалов, полученных Кеезомом путем усреднения по всем ориентациям молекул зависящих от взаимной ориентации молекул электростатических потенциалов взаимодействия всех постоянных электрических мультиполей молекул также приводит к неверным результатам для ВВК. Можно также показать, что это же имеет место и для всех вириальных коэффициентов.

Поэтому для вычисления ВВК нельзя использовать потенциал Кеезома для диполь-дипольного взаимодействия.

Следовательно, в работе [2] для описания диполь- дипольного взаимодействия полярных молекул неверно использован зависящий от температуры потенциал Кеезома вместо зависящего от относительной ориентации диполей потенциала взаимодействия твердых сфер, содержащих точечные электрические диполи.

2. Как видно из таблицы 120 на стр. 758 в [1], энергия индукции Дебая-Фалкенхагена (стр. 754 в [1]), описывающая взаимодействие постоянного электрического дипольного момента одной молекулы и моментами, индуцируемыми этим диполем в другой молекуле, может стать сравнимой с электростатическим взаимодействием электрических дипольных моментов молекул при высоких температурах. Поэтому при описании взаимодействия электрических дипольных молекул нужно учитывать не только анизотропное электростатическое взаимодействие электрических дипольных моментов молекул и дисперсионное взаимодействие (потенциал Лондона), но и энергию индукции. Энергия индукции в [2] не учтена для описания полярных молекул.

**3.** Потенциалы Лондона и Кеезома верны на больших межмолекулярных расстояниях, но они в работе [2] неверно применены для малых межмолекулярных расстояний.

**4.** Как следует из формул (3.14)-(3.17) на стр. 38 в [1], зависящий от температуры потенциал Кеезома, обратно пропорциональный шестой степени межмолекулярного расстояния, описывает только диполь-дипольное взаимодействие и взаимодействие точечного заряда с квадрупольным моментом, а не квадруполь-квадрупольное взаимодействие. В работе [2] в разделах «введение» и «теория», а также при обсуждении уравнения состояния азота и диоксида углерода (стр.695 в [2]) неверно полагают, что квадруполь-квадрупольное взаимодействие описывается этим потенциалом. Кроме того в работе [2] рассматриваются свойства нейтральных (незаряженных) молекул, еще и поэтому неверно использование в [2] потенциала Кеезома для азота и диоксида углерода.

**5.** Из уравнений состояния (7)-(10) из [2] следует, что ВВК неполярных (4) и полярных (5) молекул имеют вид

$$B(T) = v_c \cdot [c_1 - c_3(e^{-T_c/T} - 1) - c_4(e^{T_c/T} - 1) - (c_5 - c_6)T_c/T], \tag{4}$$

$$B(T) = v_c \cdot [c_1 - c_3(e^{-T_c/T} - 1) - c_4(e^{T_c/T} - 1) - (c_5 - c_6)T_c/T - (c_8 - c_9)T_c^2/T^2]. \tag{5}$$

Здесь $v_c = M\rho_c^{-1}$ - объем, приходящийся на одну молекулу (атом) в критической точке, $M$ – молярная масса молекулы, а значения безразмерных коэффициентов $c_1, c_3, ..., c_9$ указаны в таблице в [2].

Если ограничиться только первыми двумя членами разложения в ряд Тейлора экспоненты под интегралом в формуле (1), использовать суммы потенциалов Лондона и Кеезома для *r>d*, как это сделано в работе [2], то из полученной формулы для ВВК и формул (4) и (5) следует, что

$$c_5 - c_6 = (2\pi d^3 / 3v_c) \cdot (\bar{n}/d^6 kT_c), \tag{6}$$

$$c_8 - c_9 = (2\pi d^3 / 3v_c) \cdot (2(\mu^2/d^3 kT_c)^2/3), \tag{7}$$

то есть эти разности должны быть неотрицательными. Однако, как видно из таблицы в [2] разность $c_5 - c_6$ отрицательна для воды $H_2O$, что означает, что потенциал Лондона для воды является отталкивательным, что неверно. Следовательно для воды:

а) формула для ВВК не имеет молекулярного статистико-механического обоснования.

б) второй вириальный коэффициент, имеющий молекулярное статистическо-механическое основание, не является температурной функцией в уравнении состояния, вопреки утверждению [2];

в) потенциал взаимодействия (1) из работы [2] имеет неправильное поведение на далеких расстояниях, поскольку потенциал Лондона имеет притягивательный характер, а не отталкивательный.

**6.** Формула (7) неверна, если даже считать, что можно использовать потенциал Кеезома. Это легко увидеть из следующего. Ограничиваясь только первыми *тремя* членами разложения в ряд Тейлора экспоненты под интегралом в формуле для вычисления ВВК, используя суммы потенциалов Лондона и Кеезома для *r>d*, оставляя в ВВК только члены вплоть до квадратичных по обратной температуре, с помощью формул (4) и (5) получаем формулу

$$c_8 - c_9 = \left(2\pi d^3 / 3v_c\right) \cdot \left(\left(\bar{n} / d^6 kT_c\right)^2 / 6 + 2(\mu^2 / d^3 kT_c)^2 / 3\right). \tag{8}$$

Из (6) и (8) имеем

$$c_8 - c_9 = \left(c_5 - c_6\right)^2 v_c / 4\pi d^3 + 2\left(2\pi d^3 / 3v_c\right)\left(\mu^2 / d^3 kT_c\right)^2 / 3. \tag{9}$$

Если считать, что дипольный момент известен, то соотношение (9) позволяет уменьшить число подгоночных параметров на один.

**7.** Из (9) для неполярных молекул $(\mu = 0)$ имеем

$$c_8 - c_9 = \left(c_5 - c_6\right)^2 / 6(2\pi d^3 / 3v_c). \tag{10}$$

Поэтому в работе [2] для неполярных молекул нельзя было считать коэффициенты независимыми даже в случае, когда параметр потенциала Лондона подбирается.

**8.** Из таблицы из [2] видно, что разность $c_8 - c_9$ отрицательна для диоксида углерода $CO_2$. Последнее с учетом (6) и (10) означает, что параметр потенциала Лондона для диоксида углерода имеет мнимое значение, что неверно. Следовательно для диоксида углерода:

а) формула для ВВК не имеет молекулярного статистико-механического обоснования.

б) второй вириальный коэффициент, имеющий молекулярное статистическо-механическое основание, не является температурной функцией в уравнении состояния, вопреки утверждению [2];

в) потенциал взаимодействия (1) из работы [2] не существует.

**9.** Из формул в разделе «теория» в [2] легко установить соотношение

$$2\pi d^3 / 3v_c = c_1 + c_3 + c_4. \tag{11}$$

Если для азота в качестве уравнения состояния используется уравнение (9) из [2], то согласно (6), (10) и (11) с учетом значений коэффициентов из таблицы в [2] имеем

$$(c_8 - c_9) / \left(c_5 - c_6\right) = \left(c_5 - c_6\right) / 6(c_1 + c_3 + c_4) \approx 1/8,$$

из чего следует, что коэффициент $c_8$ не мог оказаться пренебрежимо малым. Причиной этого может быть то, что в [2] не была учтена связь (10) между коэффициентами. Отсюда следует, что уравнение состояния (9) из [2] для азота и формула для ВВК, полученная из этого уравнения, не имеют молекулярного статистическо-механического основания.

Неучет связи (10) может быть причиной того, что $c_8 - c_9 < 0$ для диоксида углерода.

**10.** Из формул (6) и (11) легко получить формулу

$$c = kT_c v_c^2 (c_5 - c_6)(c_1 + c_3 + c_4)(3/2\pi)^2,$$

из чего следует, что $c = 1.91 \cdot 10^{-60} эрг \cdot см^3$ для гелия, $c = 207.6 \cdot 10^{-60} эрг \cdot см^3$ для аргона и $c = 277.5 \cdot 10^{-60} эрг \cdot см^3$ для азота. Значения этого параметра, вычисленные по данным по взякости (смотрите таблицу 116 на стр. 741 в [1]) равны $c = 1.29 \cdot 10^{-60} эрг \cdot см^3$ для гелия, $c = 109.2 \cdot 10^{-60} эрг \cdot см^3$ для аргона и $c = 125.6 \cdot 10^{-60} эрг \cdot см^3$ для азота. Вычисленные из данных [2] и табличные значения параметра потенциала Лондона существенно отличаются (от 48% до 121%). Полученное из данных [2] значение $c = 1.91 \cdot 10^{-60} эрг \cdot см^3$ для гелия больше наилучшего результата Слэтера и Кирквуда $c = 1.49 \cdot 10^{-60} \, эрг \cdot см^3$ (стр. 816 [1]) на 28%. Следовательно, параметр потенциала Лондона в [2] для гелия, аргона и азота являются подгоночными и потенциал (1) из [2] для этих веществ нереалистичен.

**11.** Ограничиваясь только первыми *тремя* членами разложения в ряд Тейлора экспоненты под интегралом в формуле для вычисления ВВК, используя суммы потенциалов Лондона, энергии индукции Дебая-Фалкенхагена и анизотропного потенциала взаимодействия точечных диполей для $r > d$ и формулы (4) и (5) вместо неверных для полярных молекул формул (6) - (8) получаем формулы

$$c_5 - c_6 = \left(2\pi d^3 / 3 v_c\right) \cdot \left(c + 2\mu^2 \alpha_p\right) / d^6 k T_c,$$

$$c_8 - c_9 = \left(2\pi d^3 / 3 v_c\right) \cdot \left(\left((c + 2\mu^2 \alpha_p) / d^6 k T_c\right)^2 / 6 + (\mu^2 / d^3 k T_c)^2 / 3\right),$$

где $\alpha_p$ - поляризуемость молекулы (стр. 754-757 в [1]).

**12.** На рисунках 1 (гелий), 2 (аргон), 3 (диоксид углерода) и 4 (вода) приведена зависимость от температуры разности $\Delta B$ опытных (табличных) значений ВВК [3] от значений, вычисленных из формул (4) и (5), полученных из уравнений состояния работы [2]. Эти разности обозначены квадратиками. Зависимость от температуры разности опытных (табличных) значений ВВК от значений, вычисленных непосредственно из потенциала, полученного на основе данных [2], приведены на рисунках 1 и 2 и обозначены кружками. ВВК на этих рисунках имеет размерность $см^3 / моль$. Погрешность определения данных по ВВК в основном не превышает $\pm (1-3) см^3 / моль$. [3-5].

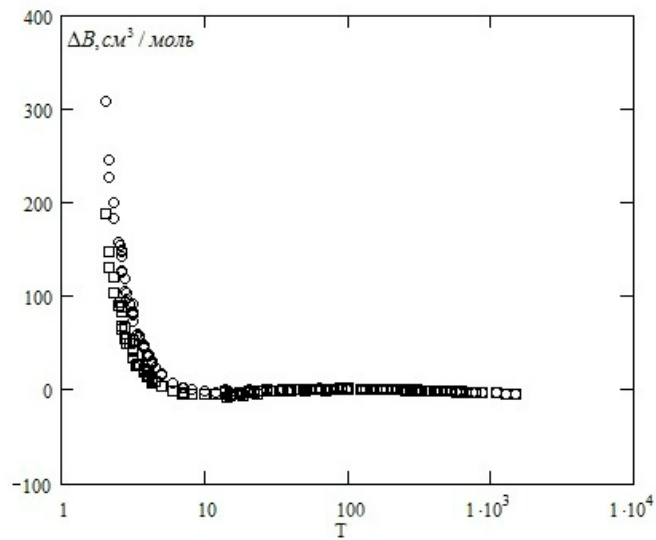

Рис. 1. Зависимости от температуры разности опытных (табличных) данных [3] по ВВК для гелия от значений, вычисленных по формуле (4) (квадратики) и разности опытных (табличных) данных [3] по ВВК от значений, вычисленных по формуле (1) из потенциала взаимодействия, восстановленного по данным [2] (кружки).

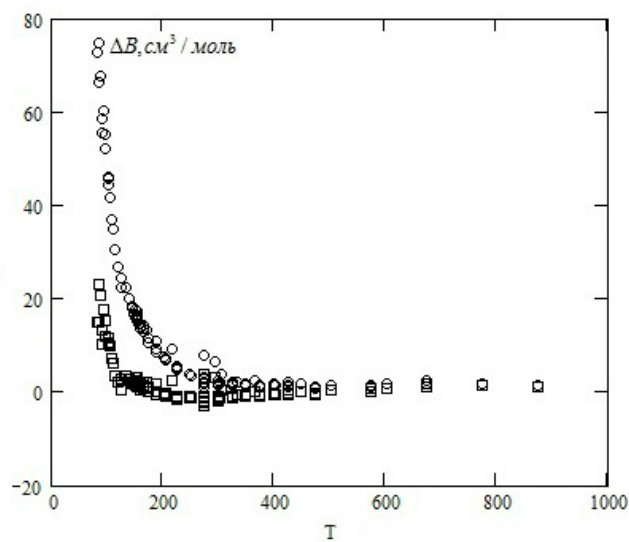

Рис. 2. Зависимости от температуры разности опытных (табличных) данных [3] по ВВК для аргона от значений, вычисленных по формуле (4) (квадратики) и разности опытных (табличных) данных [3] по ВВК от значений, вычисленных по формуле (1) из потенциала взаимодействия, восстановленного по данным [2] (кружки).

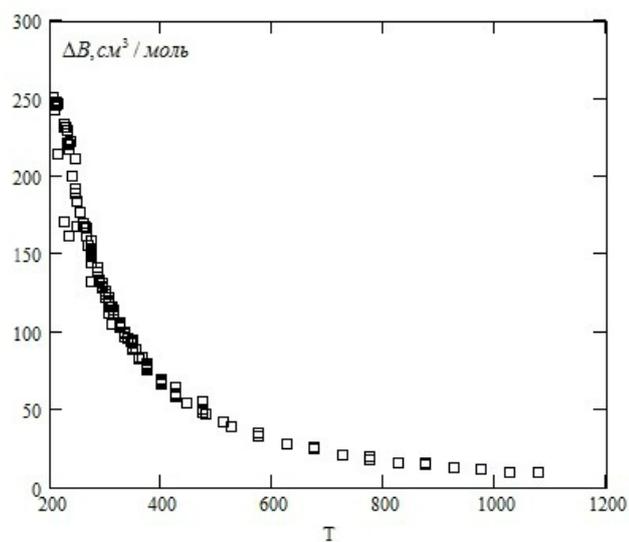

Рис. 3. Зависимость от температуры разности опытных (табличных) данных [3] по ВВК для диоксида углерода от значений, вычисленных по формуле (4) (квадратики).

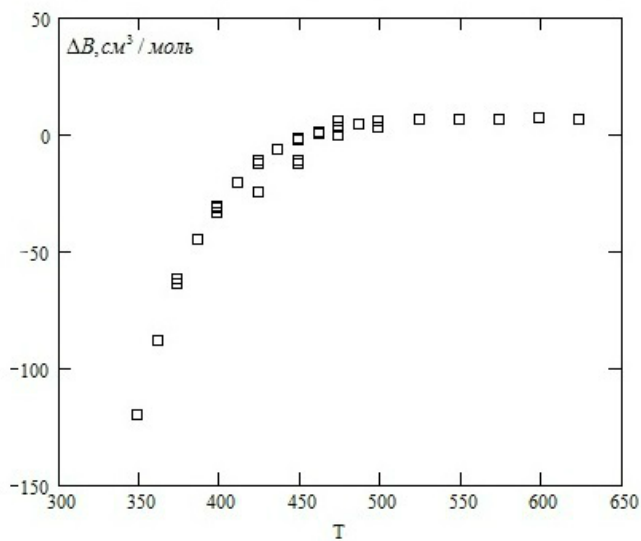

Рис. 4. Зависимость от температуры разности опытных (табличных) данных [3] по ВВК для воды от значений, вычисленных по формуле (5) (квадратики).

Как видно из рисунков 1-4, формулы (4) и (5) для ВВК, полученные из уравнений состояния из [2], не дают удовлетворительного количественного согласия с опытными (табличными) данными по ВВК.

Из рисунков 1-2 видно, что ВВК, вычисленные на основе потенциала (1) из [2], восстановленного по данным из таблицы из [2], не дают удовлетворительного количественного согласия с опытными (табличными) данными по ВВК. Видно, что эти вычисленные значения ВВК согласуются с опытными (табличными) данными хуже, чем ВВК из формулы (4).

Поэтому утверждение [2] о том, что «уравнение (7) совместно с (8) позволяет описывать термические свойства «нормальных» веществ в широком интервале параметров состояния» «с погрешностью, близкой к погрешности экспериментальных (табличных) данных», не соответствует действительности.

ОБСУЖДЕНИЕ

Табличные данные [4] для аргона получены путем экстраполяции и интерполяции на основе уравнения состояния, использованного в [4]. Оно описывает с точностью 0,2% экспериментальные данные [5], погрешность которых 0.01-0.03%. Это уравнение эмпирическое, т.е. не имеет молекулярного статистическо- механического обоснования, поскольку не выведено с помощью равновесной статистической механики исходя из потенциала взаимодействия между молекулами. Поэтому никогда нет уверенности, что экстраполяция и интерполяция с помощью этого уравнения дает правильные результаты. Кроме того, показано [6], что это уравнение имеет плохие экстраполяционные возможности. Поэтому в общем случае неверно утверждение [2], что «уравнение (7) совместно с (8) позволяет описывать термические свойства «нормальных» веществ в широком интервале параметров состояния» «с погрешностью, близкой к погрешности экспериментальных (табличных) данных». По крайней мере, это не так для аргона в отношении погрешности экспериментальных данных. Кроме того, нет уверенности, что ситуация с погрешностью описания лучше в отношении остальных веществ, рассмотренных в [2].

Уравнения состояния в [2] полностью эмпирические, так как они не имеют последовательного молекулярного статистическо-механического основания и параметры этих уравнений подгоночные. Отсюда следует, что результаты, полученные путем интерполяции и экстраполяции на основе этих уравнений, могут быть ошибочными.

После прочтения статьи [2] может возникнуть мнение, что уравнения состояния и ВВК основаны на реальных значениях параметров дисперсионного и/или дипольдипольного взаимодействия рассматриваемых молекул. Однако, как следует из изложенного в предыдущем разделе, для диоксида углерода из ВВК получается мнимое значение параметра потенциала Лондона. Для воды дисперсионное лондоновское притяжение оказывается отталкиванием. Значения параметра потенциала Лондона, полученные на основе данных [2], для гелия, аргона и азота существенно отличаются от значений этого параметра, полученных из первых принципов [1]. Отсюда следует, что такое мнение ошибочно.

В работе [2] неверно использован зависящий от температуры потенциал Кеезома для описания диполь- дипольного взаимодействия полярных молекул. Этот же потенциал также неверно использован для квадруполь- квадрупольного взаимодействия молекул азота и диоксида углерода.

Второй вириальный коэффициент, полученный из уравнений состояния [2], не дает количественного согласия с опытными (табличными) данными. Второй вириальный коэффициент, вычисленный на основе потенциала, восстановленного из данных [2], также не дает количественного согласия с опытными (табличными) данными.

Из сопоставления вышеизложенного и текста статьи [2] следует, что статья [2] внутренне не самосогласованна, содержит неустранимые внутренние противоречия и ошибки, и она представляет собой неудачную попытку молекулярного статистическо-механического обоснования эмпирических уравнения состояния, неспособных количественно описать даже второй вириальный коэффициент реальных газов.

## СПИСОК ЛИТЕРАТУРЫ